\documentclass[fleqn,usenatbib]{mnras}

\usepackage{newtxtext,newtxmath}
\usepackage{lineno}
\usepackage[T1]{fontenc}
\usepackage{indentfirst}

\DeclareRobustCommand{\VAN}[3]{#2}
\let\VANthebibliography\thebibliography
\def\thebibliography{\DeclareRobustCommand{\VAN}[3]{##3}\VANthebibliography}

\usepackage{graphicx}	
\usepackage{amsmath}	
\usepackage{mathrsfs}

\title[A model for R-X correlation in BHXRBs]{A physical model for radio and X-ray correlation in black hole X-ray binaries}

\author[Y. Jiang et al.]{
Yiheng Jiang$^{1}$, 
Shanshan Li$^{1}$, 
Xinwu Cao$^{2}$\thanks{E-mail: xwcao@zju.edu.cn (XC)},
Bei You$^{3}$\thanks{E-mail: youbei@whu.edu.cn (BY)}, 
Andrzej A. Zdziarski$^{4}$,
Saien Xu$^{3}$
\\
$^{1}$ School of Physics, Zhejiang University, 866 Yuhangtang Road, Hangzhou, 310058, People’s Republic of China\\
$^{2}$ Institute for Astronomy, School of Physics, Zhejiang University, 866 Yuhangtang Road, Hangzhou, 310058, People’s Republic of China\\
$^{3}$ Department of Astronomy, School of Physics and Technology, Wuhan University, Wuhan, 430072, People’s Republic of China\\
$^{4}$ Nicolaus Copernicus Astronomical Center, Polish Academy of Sciences, Bartycka 18, PL-00-716 Warszawa, Poland}

\date{Accepted XXX. Received YYY; in original form ZZZ}

\pubyear{2024}

\begin{document}
\label{firstpage}
\pagerange{\pageref{firstpage}--\pageref{lastpage}}
\maketitle
\begin{abstract}
A tight correlation between the radio and X-ray emission in the hard state of black hole X-ray binaries (BHXRBs) indicates an intrinsic disc-jet connection in stellar black hole accretion systems, though the detailed physics processes at work are still quite unclear. A hot accretion flow is suggested to match the outer cold thin disc at a certain radius in the hard state, which may vary with the accretion rate. In this work, we assume that the magnetic field generated in the outer thin disc is advected inwards by the inner hot accretion flow, which is substantially enhanced near the BH. Such a strong field threading the horizon of a spinning BH is responsible for launching relativistic jets in BHXRBs via the Blandford-Znajek mechanism. Thus, both the jet power and the X-ray emission increase with the mass accretion rate, and we find that our model calculations are able to reproduce the observed radio/X-ray correlation quantitatively. For some individual BHXRBs, the slopes of the radio/X-ray correlations become steeper when the sources are brighter. Our model calculations show that this feature is due to the transition of the outer disc with gas pressure dominated to radiation pressure dominated, which leads to different accretion rate dependence of the field strength in the outer disc.     
\end{abstract}


\section{Introduction}\label{sect:intro}

Black hole X-ray binaries (BHXRBs) usually transition between different states \citep{2007A&ARv..15....1D}.
It has been found that the X-ray emission is tightly correlated with the radio emission in the low/hard state of the BHXRBs \citep[][]{1998A&A...337..460H,2003MNRAS.344...60G,2003MNRAS.345.1057M,2008MNRAS.389.1697C,2014MNRAS.445..290G,2018MNRAS.478L.132G,2018MNRAS.481.4513I, doi:10.1126/science.abo4504,you2024}. If the radio emission originates in the jets in BHXRBs, a relation of the radio luminosity to the jet power is predicted based on the conical jet model \citep[][]{1979ApJ...232...34B,1995A&A...293..665F}. The X-ray emission may originate in an inefficient hot accretion flow \citep[e.g.,][]{2003MNRAS.345.1057M}. Thus, the radio/X-ray correlation observed in the low/hard state of the BHXRBs strongly indicates a close connection between the accretion and jet \citep{2004MNRAS.351..791Z}, while an alternative explanation was also suggested in the frame of the jet-dominated X-ray emission scenario \citep[][]{2003A&A...397..645M,2004A&A...414..895F}. The detailed physics at work in the disc-jet connection implied by the radio/X-ray correlation is still quite uncertain. 

A widely accepted scenario suggested that the   {different states of BHXRBs are caused by accretion mode transition}. The accretion mode transitions are mainly triggered by the variations of the mass accretion rate \citep[][]{1997ApJ...489..865E}, {which is suggested to be caused by the disc instability \citep[e.g.,][]{1993adcs.book....6C,1998MNRAS.293L..42K,1999MNRAS.303..139D,2001A&A...373..251D,2020AdSpR..66.1004H}.} {A hot accretion flow (advection dominated accretion flow; ADAF) is surrounding the BH in the low/hard state when the accretion rate is lower than a critical value, while the outer geometrically thin accretion disc approaches inner stable circular orbit (ISCO) of the BH in the thermal (soft) state of BHXRBs \citep[e.g.,][]{1996A&A...314..813L,1997ApJ...489..865E,2000A&A...354L..67M,2004MNRAS.351..791Z,2005ApJ...629..408Y,2008ApJ...682..212W,2016ApJ...817...71C,2015MNRAS.453.3447D,2016ApJ...821..104Y}.} It is theoretically proposed that, in the low/hard state, {an outer thin disc is truncated to an ADAF within a certain radius}, namely the truncated radius $R_{\rm tr}$, which increases with decreasing accretion rate, i.e., the inner ADAF expands outwards in the soft-to-hard state transition, and vice versa \citep[e.g.,][]{1996A&A...314..813L,1997ApJ...489..865E,1995ApJ...438L..37A,2014ARA&A..52..529Y}. Recently, the expansion of an ADAF with the decreasing accretion rate was observationally inferred in the BHXRB MAXI J1820+070 \citep{doi:10.1126/science.abo4504}.

{The hysteretic state transition has been observed in XRBs, i.e., the transition luminosity of the hard-to-soft state is much higher than that for the soft-to-hard state \citep[][]{1995ApJ...442L..13M,1997ApJ...477L..95Z,2002MNRAS.332..856N,2003MNRAS.338..189M,2004MNRAS.351..791Z}, which strongly implies that the accretion mode transition is not solely regulated by the accretion rate. An additional factor, the magnetic field, has been suggested to play an important role in the hysteretic state transition \citep[][]{2008ApJ...674..408B,2008MNRAS.385L..88P,2015ApJ...809..118B,2016RAA....16...48L}. \citet{2008ApJ...674..408B} suggested different values of magnetic Prandtl number in the different disc regions, {of which the transition radius is regulated by the dimensionless accretion rate}. {The different values of $\alpha$-viscosity parameter, i.e., $\alpha\sim 1$ in the ADAF, and $\alpha=0.01-0.1$ in the thin disc, are assumed in order to reproduce the hysteretic state transition} \citep[][]{2015ApJ...809..118B}. {\citet{2016ApJ...817...71C} suggested a model of magnetic outflow-driven ADAF, which shows the critical accretion rate of the ADAF varying with the large-scale magnetic field strength. \citet{2009ApJ...701.1940Y} discovered a correlation of the transition luminosity with the peak luminosity in the thermal state of X-ray binaries, which can be quantitatively reproduced with the magnetic disc-outflow model developed by} \citet{2021A&A...654A..81C}.

Steady relativistic jets are always {observed in} the low/hard state of BHXRBs, whereas the jets are usually suppressed in the high/soft state \citep[e.g.][]{   1999MNRAS.308..473F,2001ApJ...554...43C,2003MNRAS.343L..99F,2003MNRAS.344...60G,2004MNRAS.355.1105F}, which are supposed to be most probably powered by taping energy from spinning BHs with co-rotating magnetic field \citep[][]{1977MNRAS.179..433B}. The velocity of the gas magnetically driven from the disc is rather small, which may correspond to the outflows observed in BHXRBs or active galactic nuclei \citep[][]{1982MNRAS.199..883B,2019ApJ...881...34Y,2021MNRAS.505.3596D,2022ApJ...926...11L}. 

The large-scale magnetic field is a key ingredient in the jet formation mechanism, of which the origin is still quite uncertain. One of the most promising scenarios is a weak external coherent field threading the gas being advected inwards in the accretion disc \citep[e.g.,][]{1974Ap&SS..28...45B,1989ASSL..156...99V,1994MNRAS.267..235L}. However, it was argued that the advection of the field in this way is rather inefficient for a viscous thin accretion disc \citep[][]{1994MNRAS.267..235L}, because the magnetic diffusivity $\eta\sim \nu$, where $\nu$ is the turbulent viscosity \citep[][]{1979cmft.book.....P,2003A&A...411..321Y,2009A&A...507...19F,2009ApJ...697.1901G}. \citet{2013ApJ...765..149C} suggested that the external field can be substantially enhanced in a thin disc if most of its angular momentum is carried away by the magnetically driven outflows. For high mass BHXRBs, the field of the convective donor may possess a strong magnetic field at its surface, which may be advected inwards by such a thin disc with magnetic outflows even in the high/soft state \citep[][]{2020MNRAS.492..223C}. However, the field of the {the gas supplied by the} companion star is too weak to maintain such a disc-outflow system in low-mass BHXRBs, which {suppresses} the jets in high/soft state in low-mass BHXRBs \citep[][]{2019MNRAS.485.1916C}. 

The situation of the BHXRBs in the low/hard state is different from the high/soft state in that an ADAF is present in the inner region near the BH. {Thus, the high radial velocity of the ADAF makes the 
the external field significantly amplified in the inner region surrounding the BH} \citep[][]{2011ApJ...737...94C,2020MNRAS.494.4854D,2023ApJ...944..182D}. Even if the field flux supplied by the companion star is substantially amplified in the inner ADAF, it is still too weak for launching relativistic jets observed in the low/hard state of the low-mass BHXRBs \citep[see][for the detailed calculations]{2019MNRAS.485.1916C}. It implies that a coherent field significantly stronger than that of the companion star is needed for the observed jets, which should be locally generated in the outer thin disc. This coherent field is then dragged inwards by the inner ADAF, which becomes sufficiently strong to accelerate relativistic jets near the BH. There are two mechanisms that may be viable for the locally generated field in the outer thin disc.  {The inverse-cascade of the dynamo process in the thin disc \citep{1996MNRAS.281..219T} has been employed successfully in modeling the jet formation in the BHXRB MAXI J1820+070} \citep{doi:10.1126/science.abo4504}. Another scenario is based on recent numerical simulations in \citet{2020MNRAS.494.3656L}, in which the turbulence in a radially extended accretion disc is suggested to generate large-scale poloidal magnetic flux in situ under certain circumstances. However, such simulations are carried out for a thick disc, and it is found that the field generation in thin discs is not so efficient as their thick counterparts \citep[][]{2022ApJS..263...26L}.

The gas density of an ADAF is very low   {due to} its low accretion rate and high radial velocity.   {The energy transfer from the ions to electrons is very inefficient, and therefore a two-temperature accretion flow is present} \citep[][]{1994ApJ...428L..13N,1995ApJ...452..710N}.  The radiation efficiency of the ADAF is substantially lower than that of the thin disc, which decreases with decreasing accretion rate, because the radial energy advection becomes more important for a lower accretion rate \citep[see][for reviews, and the references therein]{1998tx19.confE.261N,2014ARA&A..52..529Y}. The radiation efficiency of the ADAF can be calculated based on its global structure derived with specified values of the disc parameters, though the precise values of the disc parameters have not been determined \citep[e.g.,][]{2010ApJ...716.1423X,2012MNRAS.427.1580X}.   

In this work, we propose a theoretical model for radio/X-ray correlation in the hard state of BHXRBs, which is described in Section \ref{sec:model}. The results of the model calculations and the comparison with the observations are given in Section \ref{sec:results}, and the final section contains the discussion of the results.

\section{Model}\label{sec:model}

The observations of BHXRBs suggested   {that the outer thin disc connects to the inner ADAF at the truncated radius $R_{\rm tr}$, which varies the accretion rate in the disc} \citep[][]{1997ApJ...489..865E,doi:10.1126/science.abo4504}. The hard X-ray emission is predominantly from the inner ADAF in the hard state of BHXRBs, which can be calculated with the ADAF model. 

The radio luminosity of the BHXRB is an indicator of jet power. The jets are assumed to be driven by a spinning BH with a large-scale magnetic field (i.e., the BZ mechanism), of which the power can be calculated if the field strength at the BH horizon and the value of the spin parameter $a$ are known.   {We assume that the large-scale magnetic field of the outer thin disc is generated through an inverse cascade of the field created by the dynamo} \citep[][]{1996MNRAS.281..219T,1999ApJ...512..100L,2021A&A...654A..81C}, which is then advected by the inner ADAF as an external coherent field \citep[][]{2011ApJ...737...94C,2021A&A...654A..81C}. 
The external field originating in the outer thin disc is substantially enhanced in the ADAF. The strength at the BH horizon is available as a function of the accretion rate, with which the jet power is calculated with the standard formalism of the BZ mechanism for a given BH spin $a$. 
In order to compare our model calculations with the observed radio/X-ray correlation, we have to convert the BZ jet power to the radio luminosity. In principle, the radio luminosity of the jets is not solely dependent on the jet power, which may also be affected by other physical quantities of the jets. To avoid being stuck in this issue. we estimate the radio luminosity of the jets from the jet power using an empirical relation between the jet power and radio luminosity for BHXRBs \citep[e.g.,][]{1995A&A...293..665F,doi:10.1126/science.abo4504}. Thus, a theoretical relation of the radio luminosity with the X-ray luminosity is derived by varying the value of the accretion rate.

{\subsection{ADAF with magnetically driven outflows}\label{subsec:adaf_outflows}}

In the traditional ADAF model, the critical accretion rate for an ADAF surrounding a BH is determined by the structure of the ADAF. The ADAF is suppressed when its accretion rate is above the critical value \citep[][]{1995ApJ...452..710N,1998tx19.confE.261N}.   {The structure of the ADAF with magnetic outflows is different from that of a normal ADAF if magnetic outflows carry away a fraction of the angular momentum of the gas in the ADAF. The radial velocity of such an ADAF with outflows is increased, and then the critical mass accretion rate is also increased}, which has been explored in detail by \citet{2016ApJ...817...71C}. We briefly summarize the calculations below \citep*[see][for the details]{2016ApJ...817...71C}. 

{The magnetic torque exerted by the outflows on the unit area of the ADAF is
\begin{equation}
T_{\rm m}={\frac {B_zB_{\phi}^{\rm s}}{2\pi}}R,\label{t_m_1}
\end{equation}
where $B_z$ and $B_{\phi}^{\rm s}$ are vertical and azimuthal components of the large-scale magnetic field at the disc surface, respectively. Hereafter, we use the superscript "$\prime$" to denote the quantities of the ADAF with magnetically driven outflows. The radial velocity of an ADAF with magnetic outflows is
\begin{displaymath}
V_R^\prime=V_{R}+V_{R,\rm m}=-\alpha c_{\rm s}{\frac {H}{R}}-{\frac
{2T_{\rm m}}{\Sigma
R\Omega}}=-\alpha c_{\rm s}{\frac {H}{R}}- {\frac {B_zB_{\phi}^{\rm
s}}{\pi\Sigma\Omega}}
\end{displaymath}
\begin{equation}
~~~~~~~~~~~~~~~~~~~~~~~~~~~=V_R\left(1+{\frac {B_zB_{\phi}^{\rm
s}}{\pi\Sigma\Omega}}{\frac {R}{\alpha c_{\rm
s}H}}\right), \label{v_r4}
\end{equation}
where $V_{R,\rm m}$ is the component of the radial velocity contributed by the magnetic torque, $c_{\rm s}$ is the sound speed of the gas, $\Omega$ is the angular velocity, $\Sigma$ and $H$ are the surface density and thickness of the ADAF respectively, and the $\alpha$-viscosity is adopted. It can be re-written as
\begin{equation}
V_R^\prime=V_R\left(1+{\frac {4\xi_\phi}{\tilde{H}\beta\alpha
f_\Omega}}\right)=(1+f_{\rm m})V_R,\label{v_r3}
\end{equation}
where the dimensionless quantities are defined as
\begin{equation}
\xi_\phi=-{\frac {B_{\phi}^{\rm s}}{B_z}}, ~~~\beta={\frac {p_{\rm gas}}{B_z^2/8\pi}},~~~f_{\Omega}={\frac {\Omega}{\Omega_{\rm K}}},~~~ \tilde{H}={\frac H R}, \label{def_quan}
\end{equation}
and
\begin{equation}
f_{\rm m}={\frac {4\xi_\phi}{\tilde{H}\beta\alpha f_\Omega}}.
\label{f_m}
\end{equation}
The ratio $\xi_\phi\la 1$ is required by the stability criterion \citep*[see the detailed discussion
in][]{1999ApJ...512..100L}.}

{Similar to the conventional ADAF case, the value of the critical accretion rate $\dot{M}_{\rm cr}$ for the accretion mode transition can be estimated by letting ion-electron equilibration timescale equal to the accretion timescale, which leads to
\begin{equation}
\dot{M}_{\rm cr}=(1+f_{\rm m})^2\dot{M}_{\rm cr,0}=\alpha^2(1+f_{\rm m})^2\dot{M}_{\rm Edd}, \label{mdot_crit3}
\end{equation}} where $\dot{M}_{\rm Edd}$ is the Eddington accretion rate ($\dot{M}_{\rm Edd}\equiv L_{\rm Edd}/0.1c^2=4\pi GM_{\rm BH}/0.1c\kappa_{\rm T}$), $M_{\rm BH}$ is the black hole mass, and $\dot{M}_{\rm cr,0}$ is the commonly used critical accretion rate for a normal ADAF \citep[][]{1995ApJ...452..710N,1998tx19.confE.261N}.
{For an ADAF with magnetic outflows, its luminosity can be approximated as
\begin{equation}
L_{\rm ADAF}^{\prime}\sim (1+f_{\rm m})L_{\rm ADAF}, \label{l_adaf_mag}
\end{equation}
which means the ADAF with magnetic outflows is $\sim f_{\rm m}$ times more
luminous than the conventional ADAF accreting at the critical rate. }

{The gas pressure of an ADAF with magnetic outflows is
\begin{equation}
p_{\rm gas}\simeq 1.74\times 10^{17}\alpha^{-1}(1+f_{\rm m})^{-1}m^{-1}\dot{m}_{\rm cr}r_{\rm tr}^{-5/2}\left({\frac H R}\right)^{-1}~~{\rm g~cm}^{-1}~{\rm s}^{-2}, \label{p_adaf}
\end{equation}
where
\begin{equation}
    r=\frac{Rc^2}{GM_{\rm BH}}, ~~~~~~m={\frac {M_{\rm BH}}{M_\odot}},~~~~~~\dot{m}_{\rm cr}={\frac {\dot{M}_{\rm cr}}{\dot{M}_{\rm Edd}}}.    \label{dimensionless_quan} 
\end{equation}
Combining Equations (\ref{def_quan}), (\ref{f_m}), and (\ref{p_adaf}), we derive
\begin{equation}
f_{\rm m}(1+f_{\rm m})^{-1}=9.15\times 10^{-19}m\dot{m}^{-1}_{\rm cr}r^{5/2}B_z^2
\label{f_m_2}
\end{equation}
where the approximations, $\xi_\phi\sim 1$ and $f_\Omega\sim 1$, are adopted. Thus, the factor $f_{\rm m}$, critical accretion rate $\dot{m}_{\rm cr}$, and the luminosity of the ADAF with magnetic outflows can be calculated with Equations (\ref{mdot_crit3}), (\ref{l_adaf_mag}), and (\ref{f_m_2}), while 
the values of the parameters are specified. 
\vskip 1cm

\subsection{Coherent field generation in the cold disc}\label{subsec:b-cold-disc}

The detailed physical processes responsible for the large-scale field generated in the thin disc are still quite uncertain. One of the most promising mechanisms is the inverse-cascade of the dynamo process in the thin disc as suggested by \citet{1996MNRAS.281..219T}. The size of the field loops formed with this mechanism is around the disc radius, and it is most probable that the field line threading the inner ADAF connects to the outer thin disc as a whole loop. Thus, the field line threading the ADAF can be sufficiently dragged to the BH, while it still connects the outer thin disc (or even diffuses out). A similar phenomenon has been observed in the numerical simulations \citep[e.g.,][]{2009ApJ...707..428B}. In this work, we assume the large-scale field of the thin disc is generated via the inverse-cascade of the dynamo process \citep[][]{1996MNRAS.281..219T}.   {Thus, we estimate the strength of the large-scale field of the outer disc at the truncated radius $r_{\rm tr}$ as}
\begin{equation}
B_{\rm pd}(r_{\rm tr})\sim 8.10 \times 10^{9} m^{-1/2}\dot{m}_{\rm d} r_{\rm tr}^{-7/4}~{\rm Gauss},\label{b_pd_2}
\end{equation}
when $\dot{m}_{\rm d}\ge \dot{m}_{\rm d,rg}$, where $\dot{m}_{\rm d}$ is the dimensionless mass accretion rate of the outer thin disc and
\begin{equation}
B_{\rm pd}(r_{\rm tr})\sim 2.48\times 10^{8} \alpha^{-1/20}m^{-11/20}\dot{m}_{\rm d}^{3/5} r_{\rm tr}^{-49/40}~{\rm Gauss},\label{b_pd_gas}
\end{equation}
when $\dot{m}_{\rm d}<\dot{m}_{\rm d,rg}$, where   {the threshold of the accretion rate between the radiation pressure dominant and gas pressure dominant discs} 
\begin{equation}
\dot{m}_{\rm d,rg}=1.64\times 10^{-4}\alpha^{-1/8}m^{-1/8}r_{\rm tr}^{21/16},
\label{mdot_pr}
\end{equation}
which depends very weakly on $\alpha$ or $m$ \citep[see][for the detailed calculations, and the references therein]{2021A&A...654A..81C}.

{\subsection{State transitions}\label{subsec:state-transi}}

{The accretion rate is very low in the quiescent state of BHXRBs, and the outer cold disc connects to an inner ADAF at a large radius \citep*[][]{1997ApJ...489..865E}.   {The disc becomes unstable in a certain region where the temperature of the gas approaches the hydrogen ionization temperature, which makes the annulus of the disc transit to the hot state.} The same transition occurs in the adjacent annuli since heat diffuses inwards, which triggers an outburst from the quiescent/hard state to the soft state \citep*[e.g.,][]{2002apa..book.....F}.   {The heating front finally moves inwards into the ADAF, which makes much gas being fed into the inner ADAF.} Thus, the ADAF shrinks as its accretion rate becomes higher than the critical rate at its outer edge. Finally, the ADAF is completely suppressed to a thin disc extending to the ISCO, namely, the soft state \citep*[][]{1998MNRAS.293L..42K}.   {The luminosity of the disc at this moment corresponds to the observed peak luminosity in the soft state of the BHXRB} \citep[see][for the detailed discussion]{2021A&A...654A..81C}. It means that the accretion rate $\dot{m}_{\rm d}$ can be estimated with the observed peak luminosity of the thermal state. }

In this disc instability scenario for state transitions in the BHXRBs,   {the physical processes of the hard-to-soft state transition are quite different from the soft-to-hard state transition} \citep[e.g.,][]{1998MNRAS.293L..42K}. During the hard-to-soft state transition, the accretion front of the outer cold disc accreting at $\dot{m}_{\rm d}$ moves inwards. 
As the accretion timescale of the ADAF is much shorter than that of the accretion front of the outer cold disc moving to the ISCO, the accretion rate of the ADAF roughly equals the critical rate at $R_{\rm tr}$ (see the right panel in Figure \ref{Illustrate}). 

The situation of the soft-to-hard state transition is quite different. The accretion rate of the disc decays with time, since the mass stored in the disc is gradually exhausted   {(the accretion rate in the soft state is significantly higher than the feeding rate of the gas from the donor)}. The soft-to-hard state transition occurs when an ADAF is present in the inner region of the disc (see the left panel in Figure \ref{Illustrate}). 

\vskip 1cm

\subsection{Truncated radius of the thin disc}\label{subsec:truncated-radius}

It is observationally inferred that   {the inner ADAF expands radially with decreasing accretion rate \citep[e.g.,][]{doi:10.1126/science.abo4504}, which agrees in general with the theoretical argument} \citep*[][]{1997ApJ...489..865E}, of which the detailed physics is still unclear. A power law dependence of the truncated radius with the accretion rate is predicted, whereas the different values of the power law index are expected by the calculations based on different physical scenarios \citep[][]{1995ApJ...438L..37A,1999ApJ...527L..17L,2000A&A...360.1170R,2000ApJ...538..295M,2002A&A...387..918S,2007ApJ...668.1145Q,2012ApJ...759...65T}. Thus, the relation of the truncated radius with the accretion rate is quite uncertain. In this work, to avoid being stuck in the detailed physics of this issue, we try to derive how the truncated radius varies with the accretion rate on the assumption of the so-called strong ADAF principle, i.e., an ADAF is always present whenever $\dot{m}(r)<\dot{m}_{\rm cr}(r)$ \citep[][]{1998tx19.confE.261N}.  

  {The critical accretion rate for an ADAF can be estimated by equating the accretion timescale with the Coulomb interaction timescale between electrons and ions}, which reads \citep[see Equation 3.17 in][]{1998tbha.conf..148N}
\begin{equation}
    \dot{m}_{\rm cr}(r) \propto T^{3/2}_{\rm e}(r)\alpha^2, \label{mdot_te}
\end{equation}
where the electron temperature $T_{\rm e}$ varies with radius as a power law in the outer region of the ADAF, while the temperature distribution is rather flat in the inner region of the ADAF near the BH \citep[e.g.,][]{2000ApJ...534..734M,2010ApJ...716.1423X}. The most commonly used critical accretion rate for an ADAF without magnetic outflows surrounding a BH is \citep[][]{1995ApJ...452..710N,1998tx19.confE.261N}
\begin{equation}
    \dot{m}_{\rm cr,0}\equiv  {\frac {\dot{M}_{\rm cr,0}}{\dot{M}_{\rm Edd}}} \sim\alpha^2, \label{mdot_cr}
\end{equation}
in which the size of the ADAF is implicitly assumed to be small, and the electron temperature is, therefore, roughly constant. The ADAF will be completely suppressed when the accretion rate $\dot{m}\ga \dot{m}_{\rm cr,0}$.   {It implies that the critical accretion rate of an ADAF decreases with increasing ADAF size}, because the electron temperature always declines with increasing radius in an ADAF. The critical accretion rate of the ADAF truncated at $r_{\rm tr}$ is 
\begin{equation}
  \dot{m}_{\rm cr}(r_{\rm tr})= \dot{m}_{\rm cr,0}\left({\frac {r_{\rm tr}}{r_{\rm tr,0}}} \right)^{-3\alpha_{\rm e}/2}\simeq \left({\frac {r_{\rm tr}}{r_{\rm tr,0}}} \right)^{-3\alpha_{\rm e}/2}\alpha^2, \label{mdot_cr_tr}
\end{equation}
where Equations (\ref{mdot_te}) and (\ref{mdot_cr}) are used, and $T_{\rm e} \propto r^{-\alpha_{\rm e}}$ is assumed, which is a good approximation if its radius is not very close to $r_{\rm ISCO}$ \citep[e.g.,][]{2000ApJ...534..734M,2010ApJ...716.1423X}. 

Thus we derive the relation between $r_{\rm tr}$ and $\dot{m}_{\rm cr}$ for an ADAF without magnetic outflows as 
\begin{equation}
    r_{\rm tr} = r_{\rm tr,0}\left[\frac{\dot{m}_{\rm cr}(r_{\rm tr})}{\dot{m}_{\rm cr}(r_{\rm tr,0})}\right]^{{-2}/{3 \alpha_{\rm e}}}.
    \label{htsrtr}
\end{equation}
A relation of $r_{\rm tr}$ with $\dot{m}$ derived in a similar way was tentatively confronted with the observations, which seems to be qualitatively consistent \citep[see Figure 8 in][]{1998tbha.conf..148N}. Base on Equation (\ref{htsrtr}), the critical accretion rate of an ADAF with magnetic outflows at $R_{\rm tr}$ can be calculated with Equations (\ref{mdot_crit3}), (\ref{l_adaf_mag}), and (\ref{f_m_2}), in which the generated field strength in thin disc $B_{\rm pd}(r_{\rm tr})$ is calculated by letting $\dot{m}=\dot{m}_{\rm d}$ in Equation (\ref{b_pd_2}) or (\ref{b_pd_gas}). {The relations between the critical accretion rate of the ADAF with magnetic outflows and the truncated radius, during the state transitions, are available from the calculations described in Section \ref{subsec:adaf_outflows}.

\vskip 1cm

\subsection{X-ray luminosity $L_{\rm X}$}\label{subsect:l_x}

The radiation efficiency of a standard thin disc surrounding a BH with dimensionless spin parameter $a$ is
\begin{equation}
\eta_{\rm rad}^{\rm sd}(a)=1-\widetilde{E}_{\rm ISCO}=1-\frac{r_{\rm ISCO}^2-2r_{\rm ISCO}+a\sqrt{r_{\rm ISCO}}}{r_{\rm ISCO}(r_{\rm ISCO}^2-3r_{\rm ISCO}+2a\sqrt{r_{\rm ISCO}})^{1/2}},
\end{equation}
where 
\begin{equation}
r_{\rm ISCO}= r_{\rm ms}= 3+Z_2-[(3-Z_1)(3+Z_1+2Z_2)]^{1/2},   
\end{equation}
when $a>0$, $Z_1$, and $Z_2$ are
\begin{equation}
    Z_1 \equiv 1+\left (1-{a^2}\right )^{1/3}\left [\left (1+{a}\right )^{1/3}+\left (1-{a}\right )^{1/3}\right ],
\end{equation}

\begin{equation}
    Z_2 \equiv \left (3{a^2}+Z_1^2\right )^{1/2},
\end{equation}
respectively \citep*[][]{bardeen1972rotating}.

The radiation efficiency of an ADAF $\eta_{\rm ADAF}\sim\eta_{\rm rad}^{\rm sd}$, when its accretion rate $\dot{m}\sim \dot{m}_{\rm cr}$. It decreases rapidly with decreasing mass accretion rate $\dot{m}$, which can be calculated with the ADAF model if the values of the parameters are specified \citep[e.g.,][]{2010ApJ...716.1423X,2012MNRAS.427.1580X}. It is found that the radiation efficiency of an ADAF can be fairly well described by 
\begin{equation}
\eta_{\rm ADAF}=\left({\frac {\dot{m}} {\dot{m}_{\rm cr}}}\right)^{s}\eta_{\rm rad}^{\rm sd}, \label{eta_adaf}    
\end{equation}
where 
\begin{equation}
 \dot{m}_{\rm cr}=(1+f_{\rm m})^2\dot{m}_{\rm cr,0}=\alpha^2(1+f_{\rm m})^2.  \label{mdot_crit4}   
\end{equation}
where Equation (\ref{mdot_cr}) for a conventional ADAF without magnetic outflows is used. The value  
of the parameter $s=0.2-1.1$ is suggested for an ADAF surrounding a Kerr BH for different values of the ADAF parameter \citep[see][for the details]{2010ApJ...716.1423X}. It was argued that outflows are inevitably driven from hot accretion flows \citep[][]{1994ApJ...428L..13N}, whereas the mass loss rate of the outflows has not been well constrained either by the observations or theoretical works, which is usually assumed to be described in a parameterized way \citep[][]{1999MNRAS.303L...1B}. Recently, the hard X-ray region of MAXI J1820+070 was observationally inferred to be outflowing via the X-ray spectral fits \citep{2021NatCo..12.1025Y}.   In this work, we consider an ADAF with magnetically driven outflows, which is different from a conventional ADAF with hydrodynamic outflows. So far as we know, a self-consistent solution to an ADAF with magnetic outflows is still unavailable. In our present work, we focus on the radiation efficiency of the ADAF, which suffers from the degeneracy of mass loss rate in the outflows and the fraction of gravitational power $\delta$ directly heating the electrons in the ADAF \citep[see][and the references therein]{2014ARA&A..52..529Y}.  \citet{2022ApJ...926...11L} explored the dynamics of magnetically driven outflows from the hot corona above the disc and found only a small fraction of the gas in the disc has been driven into the outflows.   {The corona is somewhat akin to the ADAF, so we neglect the mass loss in the magnetic outflows from the ADAF}. We assume that the uncertainties arising from the value of $\delta$ and the mass loss rate in the outflows are included in the parameter $s$ in Equation (\ref{eta_adaf}).

The hard X-ray emission is predominantly from the inner ADAF in the hard state of BHXRBs, so the hard X-ray luminosity is 
\begin{equation}
 L_{\rm X}\simeq \eta_{\rm ADAF}\dot{m}\dot{M}_{\rm Edd}c^2. \label{l_x}   
\end{equation}

\subsection{Radio luminosity $L_{\rm R}$}
\label{subsec:l_r}
 
In the hard state of low-mass BHXRBs, the field generated in the outer thin disc is taken as an external field for the inner ADAF, i.e., $B_{\rm z}(r_{\rm tr})=B_{\rm pd}(r_{\rm tr})$, which is dragged inwards and enhanced in the ADAF. In principle, we can calculate the magnetic field advection/diffusion based on a global solution of the ADAF with magnetically driven outflows by using the approach given in the previous work \citep[][]{2011ApJ...737...94C}, when the value of the magnetic Prandtl number $P_{\rm m}\equiv \eta/\nu$ ($\eta$ is the magnetic diffusivity and $\nu$ is the turbulent viscosity) is specified. The dynamics of the ADAF can be described by the self-similar solution quite well \citep[][]{1994ApJ...428L..13N}.  The relative vertical thickness $H/R$ of the ADAF is around unity. Thus, for simplicity, we adopt the self-similar solution of the ADAF to calculate the field advection/diffusion in the inner ADAF. The value of $H/R$ is taken as an input model parameter in the calculation of $v_{R}$ of the inner ADAF between $r_{\rm ISCO}$ and $r_{\rm tr}$, 

\begin{equation}
v_{R}=-{\frac {\alpha c_{\rm s} H(1+f_{\rm m})}R}=-{\frac 1 2}\alpha v_{\rm K}(R)(1+f_{\rm m})\left({\frac H R}\right)^2,
\label{v_r2}
\end{equation}
where $v_{\rm K}$ is the Keplerian velocity, and the relative thickness of the ADAF, $H/R\simeq2c_{\rm s}/v_{\rm K}$, has been used. The field advection/diffusion in the ADAF can be calculated when the value of the Prandtl number is specified \citep[see][for the details]{2011ApJ...737...94C}, and the external coherent field strength is given by Equations (\ref{b_pd_2}) or (\ref{b_pd_gas}).The field advection in the ADAF for a given radial velocity distribution can be calculated by solving the induction equation \citep[see Equation 29 in][]{2011ApJ...737...94C}, when the values of the model parameters $\alpha$, and $P_{\rm m}$, are specified. The field strength $B_{\rm h}$ at the BH horizon is available for a given value of $\dot{m}$. 

We estimate the jet power with the formula,

\begin{equation}
    P_{\rm jet} = \frac{c}{96 \pi^2 r^2_{\rm g}} \Phi^2_{\rm BH} \omega^2_H = \frac{1}{24} B^2_{\rm h} R^2_{\rm h} c a^2,
    \label{p_jetmax}
\end{equation}
where

\begin{equation}
    \Phi_{\rm BH}=2 \pi B_{\rm h} R^2_{\rm h},
\end{equation}
\begin{equation}
    \omega_{\rm H}=\frac{a}{1+\sqrt{1-a^2}},
\end{equation}
and
\begin{equation}
    R_{\rm h}=\frac{GM_{\rm BH}}{c^2}(1+\sqrt{1-a^2}),     \label{r_h}
\end{equation}
is the radius of the BH horizon \citep[][]{2015ASSL..414...45T}.

The relation of the radio luminosity of the jets with the jet power is crucial in our model, which is somewhat model dependent, and it may even vary with the model parameters \citep[e.g.,][]{1995A&A...293..665F,2003A&A...397..645M,2022ApJ...925..189Z}. For our present investigation on the statistical correlation between radio and X-ray emission for a sample of BHXRBs, we believe that the radio luminosity is the best available indicator of the jet power. For simplicity, we adopt a relation widely used   
for BHXRBs, 

\begin{equation}
	L_{\rm R}=L_{\rm 0}\left(\frac{P_{\rm jet}}{ P_{\rm 0}}\right)^{\frac{17}{12}-\frac{2\zeta}{3}}, \label{l_r}
\end{equation}

to convert the jet power to radio luminosity in our model calculations \citep[e.g.,][]{1995A&A...293..665F,2003A&A...397..645M,doi:10.1126/science.abo4504}, where $\zeta$ is radio spectral index for BHXRBs. The values of parameters, $P_{\rm 0} \sim 10^{37} - 10^{39}$ ${\rm erg \enspace s^{-1}}$ and $ L_{\rm 0} = 4.53\times 10^{30}$${\rm erg \enspace s^{-1}}$ and $1.34\times 10^{30} $${\rm erg \enspace s^{-1}}$ (correspond to the radio luminosity at 15GHz of MAXI J1820+070 at 58220 and 58397) are adopted as suggested by
\citet{2022ApJ...925..189Z}.  Throughout this work, we adopt $P_{\rm 0} = 10^{37}{\rm erg \enspace s}^{-1}$, and $L_{\rm 0} = 4.53\times 10^{30} {\rm erg \enspace s^{-1}}$ to convert jet power to radio luminosity of the jets with Equation (\ref{l_r}). In most of the calculations of the radio luminosity, we adopt $\zeta=0$, while the values other than zero are also adopted for comparison.

\section{results}\label{sec:results}

  {The accretion rate of the ADAF is lower than a critical rate $\dot{m}_{\rm cr,0}$ in the hard state of BHXRBs.} The theoretical model calculations show that the critical rate $\dot{m}_{\rm cr,0}\propto \alpha^2$ \citep[][]{1998tx19.confE.261N}. There is strong observational evidence for the soft-to-hard state transition in most BHXRBs occurring at $L_{\rm bol}/L_{\rm Edd}\sim 0.01$ \citep[e.g.,][]{2008ApJ...682..212W}, roughly corresponding to $\alpha\sim 0.1$ \citep[][]{1997ApJ...489..865E,1998tx19.confE.261N}, though it may vary for some individual sources, which is roughly consistent with the MHD numerical simulations \citep[e.g.,][]{1998ApJ...501L.189A,2002ApJ...573..738H}. We adopt $\alpha=0.1$ in most calculations of this work.

  {We adopt a typical value of $H/R=1$ in the calculations of the field advection in the ADAF.} For isotropic turbulence, the magnetic Prandtl number $P_{\rm m}\sim 1$ is expected \citep[][]{1979cmft.book.....P}. Numerical simulations indicate $P_{\rm m}\sim 1$ \citep[][]{2003A&A...411..321Y,2009A&A...507...19F,2009ApJ...697.1901G}, or $P_{\rm m}\sim 0.2-0.5$ \citep*[][]{2009A&A...504..309L}.  In this work, we adopt $P_{\rm m}=0.2-0.5$ in modeling radio/X-ray correlations.

The truncated radius $r_{\rm tr,0}$ of the outer thin disc accreting at $\dot{m}_{\rm cr,0}$ is an input model parameter. It should be very near the BH, though the precise value is still not strictly constrained by the observations \citep[][]{2014ARA&A..52..529Y}. In this work, we take $r_{\rm tr,0}=20$ and $\dot{m}_{\rm cr,0} = \alpha^2$ in all of our model calculations, and then the truncated radius $r_{\rm tr}$ as a function of $\dot{m}$ is available with Equation (\ref{htsrtr}). The final results are insensitive to the exact value of $r_{\rm tr,0}$ adopted. The external field is presumed to be generated in the outer thin disc, the strength of which is estimated with Equation (\ref{b_pd_2}) or (\ref{b_pd_gas}) depending on $\dot{m}$. This external field is further advected inwards by the inner ADAF, which is calculated with the same approach in \citet{2011ApJ...737...94C}.

{We note that the field advection in the ADAF is available for a given radial velocity distribution of the ADAF. {The radial velocity of an ADAF with magnetic outflows is amplified by a factor of $f_{\rm m}$} compared with a conventional ADAF (see Equation \ref{v_r2}). The value of $f_{\rm m}$ is mainly determined by the magnetic field strength/configuration (see Equation \ref{f_m_2}).   {It means that the ADAF structure is impacted by the field strength and configuration. On the other hand, the field properties are also determined by the ADAF. The ADAF is, therefore, closely coupled with the magnetic field. In this work, we first calculate the field strength/configuation of a normal ADAF without magnetic fields and then calculate the magnetic torque exerted on the ADAF, thus the structure of the ADAF is derived.} Based on this newly obtained ADAF structure, we re-calculate its field. We iterate such calculations till the derived ADAF and field structures converge.   {The field advected in the ADAF with magnetically driven outflows is finally obtained} (see section \ref{subsec:l_r}). We plot the amplified field strengths as functions of radius in Figure \ref{fig:1}, and it is found that the external field is significantly enhanced in the inner ADAF, which is able to drive relativistic jets near a rotating BH.} Our calculations of the field advection in the ADAF are carried out in the Newtonian frame, while the radius of the BH horizon is given by Equation (\ref{r_h}). We believe that it is a fairly good approximation of the calculation in the general relativistic frame, which is beyond the scope of this work.      

With the derived field strength at the BH horizon $B_{\rm h}$, the BZ jet power and then its radio luminosity varying with $\dot{m}$ are calculated with Equations (\ref{p_jetmax}) and (\ref{l_r}). In most of the model calculations, we assume the spectral index $\zeta = 0$. However, a few well-observed sources show different values of the spectral index $\zeta$ and rapid changes, e.g., the spectral index of GX339-4 is sometimes around $+0.25$, and for MAXI J1348-630, its index varies between $0$ and $+0.27$ in hard state\citep[][]{2013MNRAS.428.2500C,2021MNRAS.504..444C}{}{}. Thus, in Figure \ref{fig:3} we take $\zeta$ ranging from $0$ to $0.25$ with spin $a=0.99$, and from $-0.5$ to $0$ with spin $a=0.9$ to show how the results vary with $\zeta$ (see Equation \ref{l_r}).  Using Equation (\ref{l_x}), we estimate the X-ray luminosity of the BHXRB in the hard state varying with the dimensionless accretion rate $\dot{m}$, provided the value of $s$ is specified. The BH mass $m=10$ is adopted in our model calculations. Thus, we repeat the calculations of $L_{\rm X}$ and $L_{\rm R}$ by varying the values of $\dot{m}$, and finally a theoretical relation of the X-ray luminosity to the radio luminosity is derived with our model. There are several parameters, $\alpha_{\rm e}$, $s$, and $\alpha$, in our model. 

For the soft-to-hard state transition, the mass accretion rate of the inner ADAF at $r_{\rm tr}$ is the same as that of the outer thin disc (see the discussion in Section \ref{subsec:state-transi}, and the left panel in Figure \ref{Illustrate}). The comparisons of the model calculations with the observed radio/X-ray correlation are plotted in Figures \ref{fig:2} and \ref{fig:3}.
We find that previously established correlation of $L_{\rm R}\propto L_{\rm X}^{0.6}$ can be reproduced by our model calculations quite well with suitable values of the parameters. We note that only the data points obtained in soft-to-hard state transitions are included in Figures \ref{fig:2} and \ref{fig:3}, which are slightly different from those reported in the previous works.    

It is well known the transition luminosity of the hard-to-soft state is much higher than that of the soft-to-hard state, namely, the hysteretic state transition. The magnetic field may play an important role in hysteretic state transition (see the discussion in Section \ref{subsec:state-transi}). The observed correlation of the peak luminosity of the soft state with the transition luminosity in the hard-to-soft state transition can be reproduced by the model of an ADAF with magnetic outflows \citep[][]{2021A&A...654A..81C}. In this model, the accretion rate of the outer thin disc $\dot{m}_{\rm d}$ can be derived with the peak luminosity in the soft state. Based on this model, we calculate the $L_{\rm X}-L_{\rm R}$ relation for the hard-to-soft state transition by varying the value of $\dot{m}_{\rm d}$ (see the discussion in Section \ref{subsec:state-transi}). The calculations are almost the same as those aforementioned for soft-to-hard state transition, but letting $\dot{m}=\dot{m}_{\rm d}$ in Equations (\ref{b_pd_2}) and (\ref{b_pd_gas}) while calculating the external field strength $B_{\rm pd}$. 

During the hard-to-soft state transition, the inner ADAF is regulated by the critical rate $\dot{m}_{\rm cr}$ at $r_{\rm tr}$ (Equation \ref{mdot_crit4}), i.e., $\dot{m}\simeq \dot{m}_{\rm cr}(r_{\rm tr})$. The truncated radius $r_{\rm tr}$ decreases with time, and therefore the accretion rate of the inner ADAF varies with $r_{\rm tr}$. We find that the correlation of $L_{\rm X}-L_{\rm R}$ in the hard-to-soft state transition is much steeper than the normal branch of the radio/X-ray correlation with $\beta\simeq 0.6$ when $\dot{m}_{\rm d}$ is so high that the thin disc is radiation pressure dominant at $r_{\rm tr}$. Such a steeper radio/X-ray correlation has been found in some individual BHXRBs when the source is brighter (mostly during the hard-to-soft state transition), namely, a hybrid radio/X-ray correlation \citep[][]{2011MNRAS.414..677C,2020MNRAS.491L..29W,2021MNRAS.505L..58C}.  The physical origin of such a hybrid radio/X-ray correlation is still highly debated, for which a variety of different scenarios have been proposed \citep[see][for a summary]{2021MNRAS.505L..58C}, consisting of those based on radio-quiet hypothesis \citep[e.g.,][]{2009ApJ...699.1919P,2011MNRAS.414..677C,2018MNRAS.478.5159M,2019ApJ...871...26K,2019Natur.569..374M,2021MNRAS.504..444C}, or the models in the frame of the X-ray-loud hypothesis \citep[e.g.,][]{2004MNRAS.354..953Y,2011MNRAS.414..677C,2012MNRAS.427.1580X,2014MNRAS.440..965H,2014A&A...562A.142M,2016MNRAS.456.4377X}.

In our model, a rather different slope of the radio/X-ray correlation is expected for the hard-to-soft state transition due to its relatively higher accretion rate than that in the soft-to-hard state transition. We found that the data have not been categorized into two populations in almost all previous investigations on the hybrid correlations. Furthermore, as the transition luminosity varies in a large range for different sources \citep[][]{2009ApJ...701.1940Y}, we have to model the hybrid radio X-ray correlations for individual sources. We search the literature, and tentatively apply our model calculations to the observations of BHXRBs H1743$-$332 and MAXI J1348$-$630, which indeed exhibit different slopes of the radio/X-ray correlations in the soft-to-hard and hard-to-soft state transitions. In Figure \ref{fig:5} and \ref{fig:6}, we compare our model calculations with the hybrid correlations observed in these two sources. It is found that the observed hybrid correlations can be reproduced by our model calculations fairly well.

\begin{figure*}
	\includegraphics[width=\textwidth]{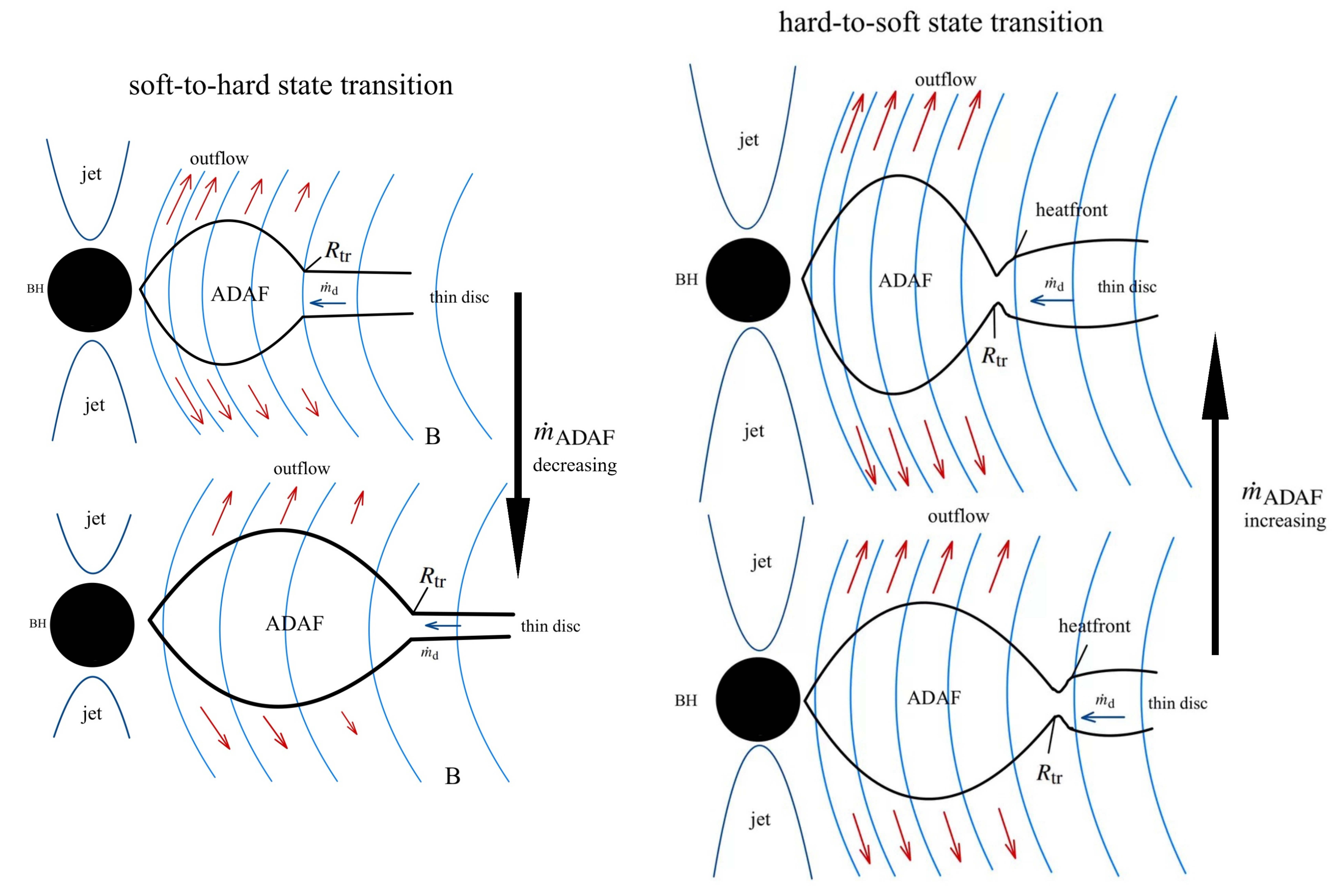}
	\caption{Illustration of the model, the strong large-scale magnetic field is formed through accumulated coherent external field in outer thin disc. ADAF is truncated to a thin disc at $R_{\rm tr}$ that decreases with increasing accretion rate of ADAF $\dot{m}_{\rm ADAF}$. The left panel:  soft-to-hard state transition, while the right panel: hard-to-soft state transition.}
	\label{Illustrate}
\end{figure*}

\begin{figure}
	
	\includegraphics[width=\columnwidth]{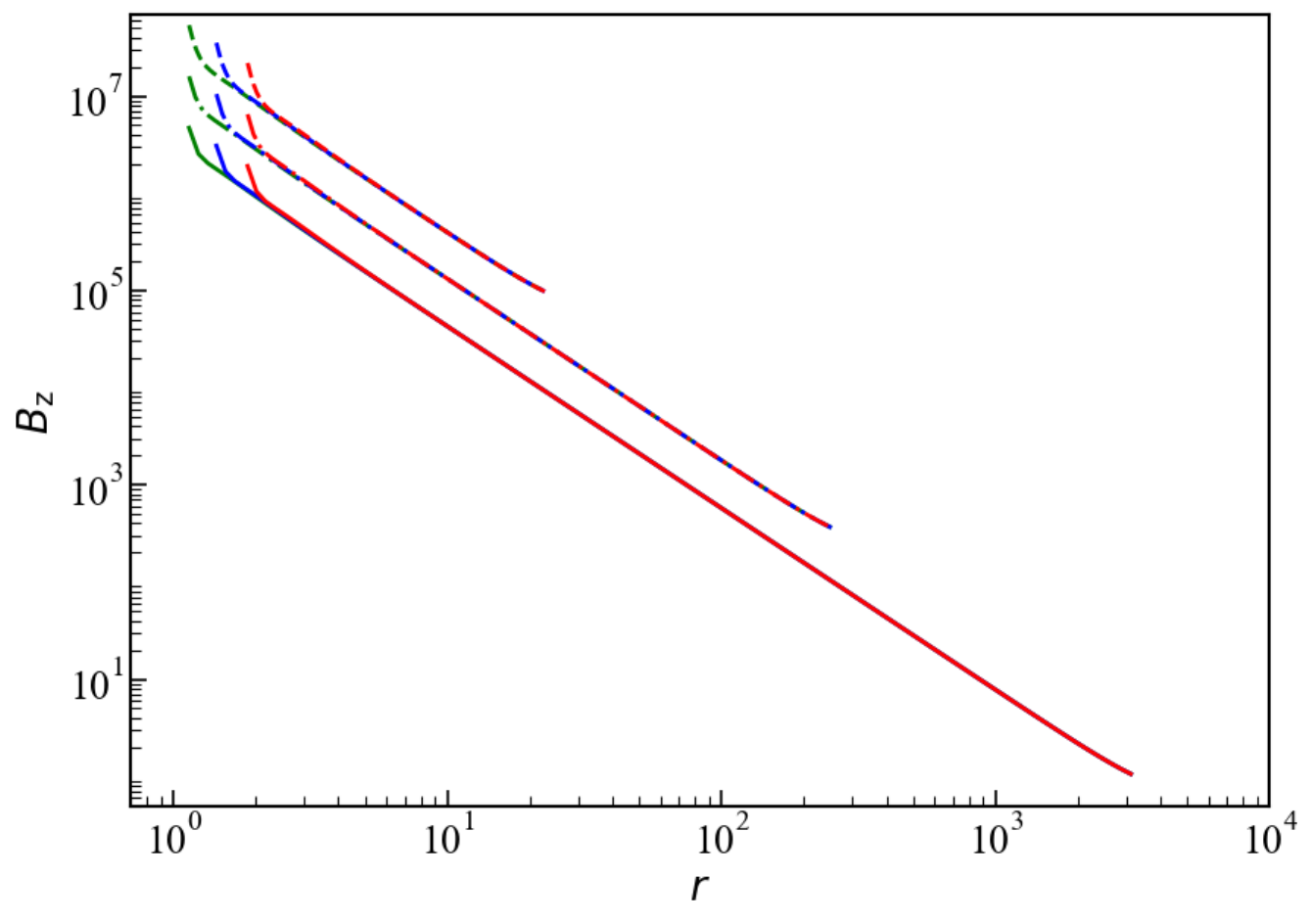}
	\caption{The vertical magnetic field strength in the accretion flow varies with radius in the soft-to-hard state transition. 
 The magnetic Prandtl number $P_{\rm m}=0.2$ is adopted in the calculations.  
 The lines with different colours correspond to different values of $a = 0.5$ (red), $a=0.9$ (blue), and $a=0.99$ (green). The dashed, solid, and dash-dotted lines represent the results of accretion rates $\Dot{m} = 1.0 \times 10^{-2}$, $1.0 \times 10^{-4}$ and $1.0 \times 10^{-6}$, respectively. The viscosity parameter $\alpha=0.1$
 is adopted in our calculations. }
	\label{fig:1}
\end{figure}

\begin{figure*}

        \centering
	\includegraphics[width=\textwidth]{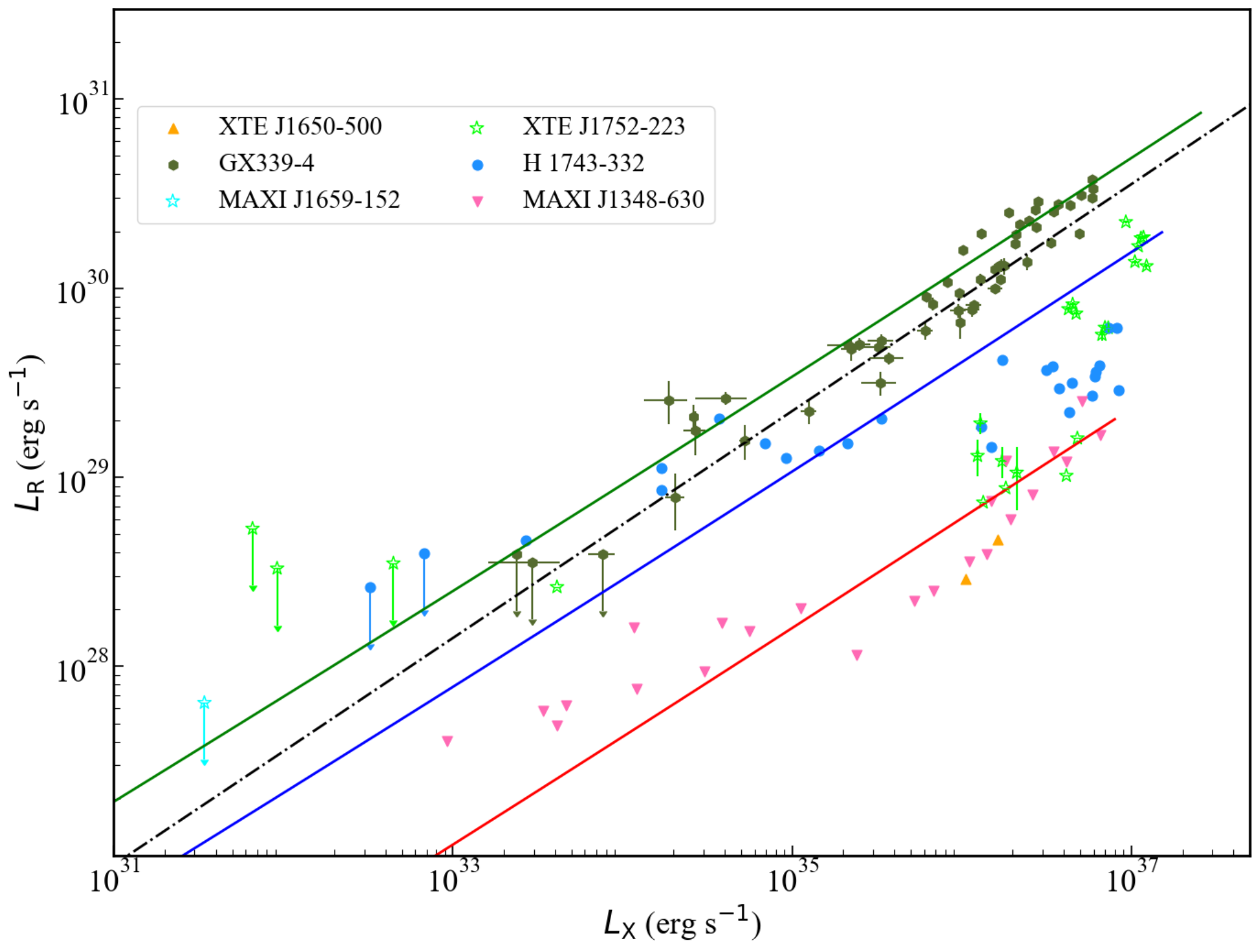}

	\caption{The non-linear relation between Radio luminosity $L_{\rm R}$ and X-ray luminosity $L_{\rm X}$ in BHXRBs ($L_{\rm R} \propto L_{\rm X}^{\beta}$). The data points are taken from the literature fitted with a regression slope $\beta = 0.6$ (dashed-dotted black line). The model calculations are carried out with the parameter values: $P_{\rm m}=0.2$, $\alpha_{\rm e} = 0.83$, $s = 0.3$, and  $\alpha=0.1$ are plotted with the solid lines. The different colour lines correspond to the results calculated with different values of the BH spin parameters $a = 0.5$ (red), $a = 0.9$ (blue), and $a = 0.99$ (green). }
	\label{fig:2}
\end{figure*} 

\begin{figure*}
	\includegraphics[width=\textwidth]{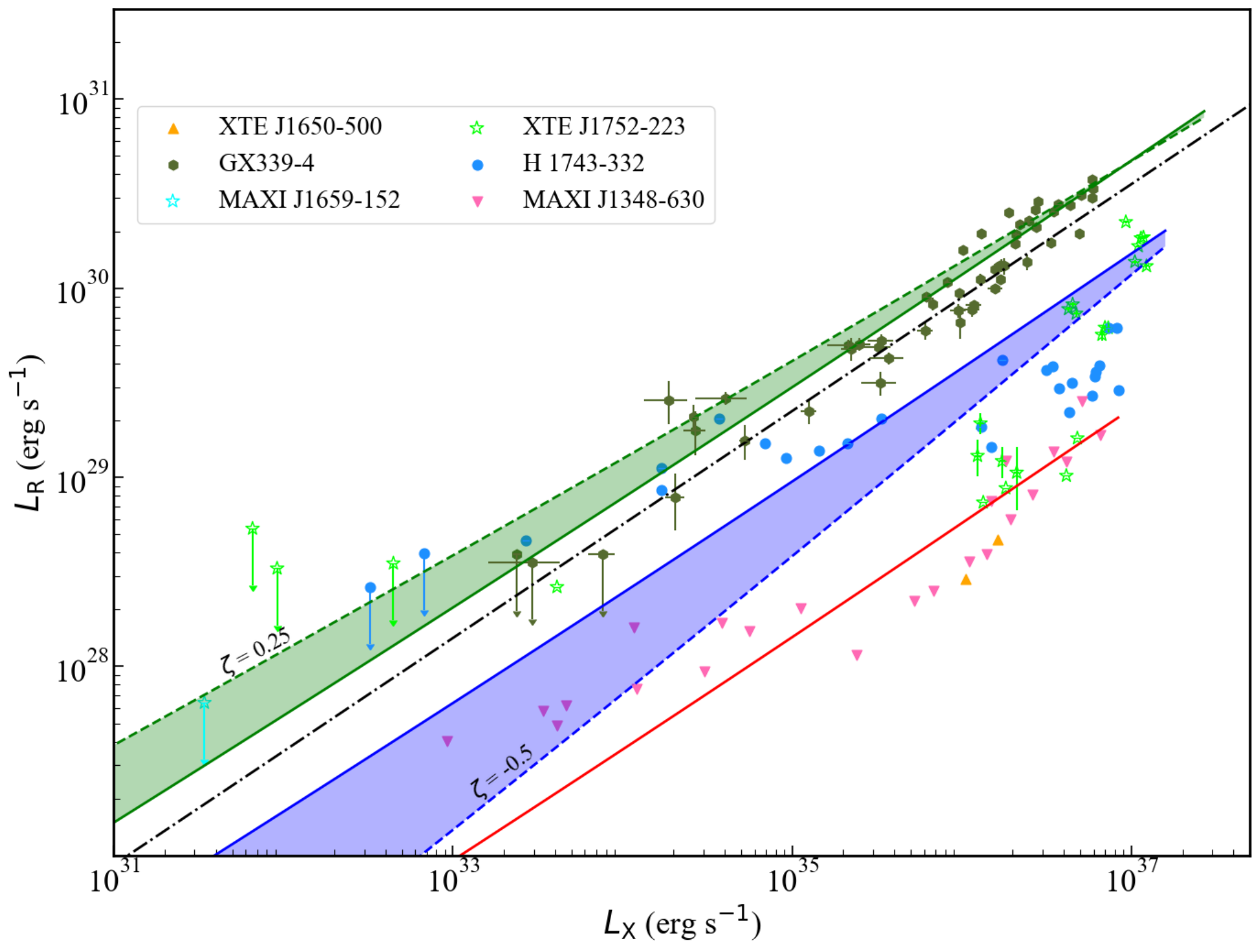}
	\caption{The same as Figure \ref{fig:2}, but for the different values of the parameters: $\alpha_{\rm e}=1$, and $s=0.6$. The spectral index $\zeta$ ranging from $0$ to $0.25$ with $a=0.99$ (green area), while from $-0.5$ to $0$ with $a=0.9$ (blue area) is adopted. }
	\label{fig:3}
\end{figure*}

\begin{figure}
	\includegraphics[width=\columnwidth]{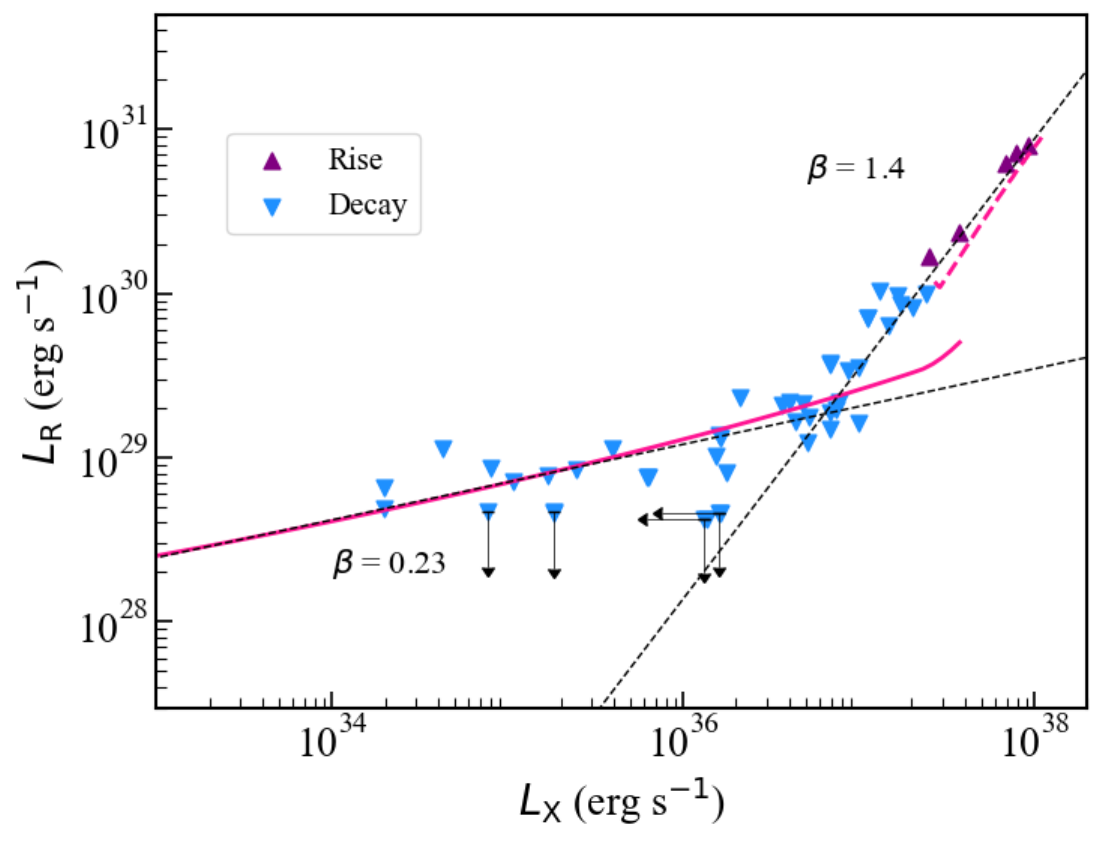}
    	\caption{The data points of H1743-332 both in the hard-to-soft and soft-to-hard state transitions. The triangles pointing upwards and downwards represent the different state transitions, i.e., hard-to-soft state transition (blue) and soft-to-hard state transition (purple), respectively. The slopes of the regression lines for the soft-to-hard and hard-to-soft state transitions are $0.23$ and $1.4$ respectively (dashed black lines). The solid pink line represents a soft-to-hard state transition, while the dashed pink line represents a hard-to-soft state transition. The model calculations are carried out with $P_{\rm m}=0.5$, 
      $\alpha_{\rm e} = 0.37 $  $\alpha = 0.16$, $s=1$, $\dot{m}_{\rm d} = 0.03$, and $a=0.87$. }
	\label{fig:5}
\end{figure}

\begin{figure}
    
    \includegraphics[width=\columnwidth]{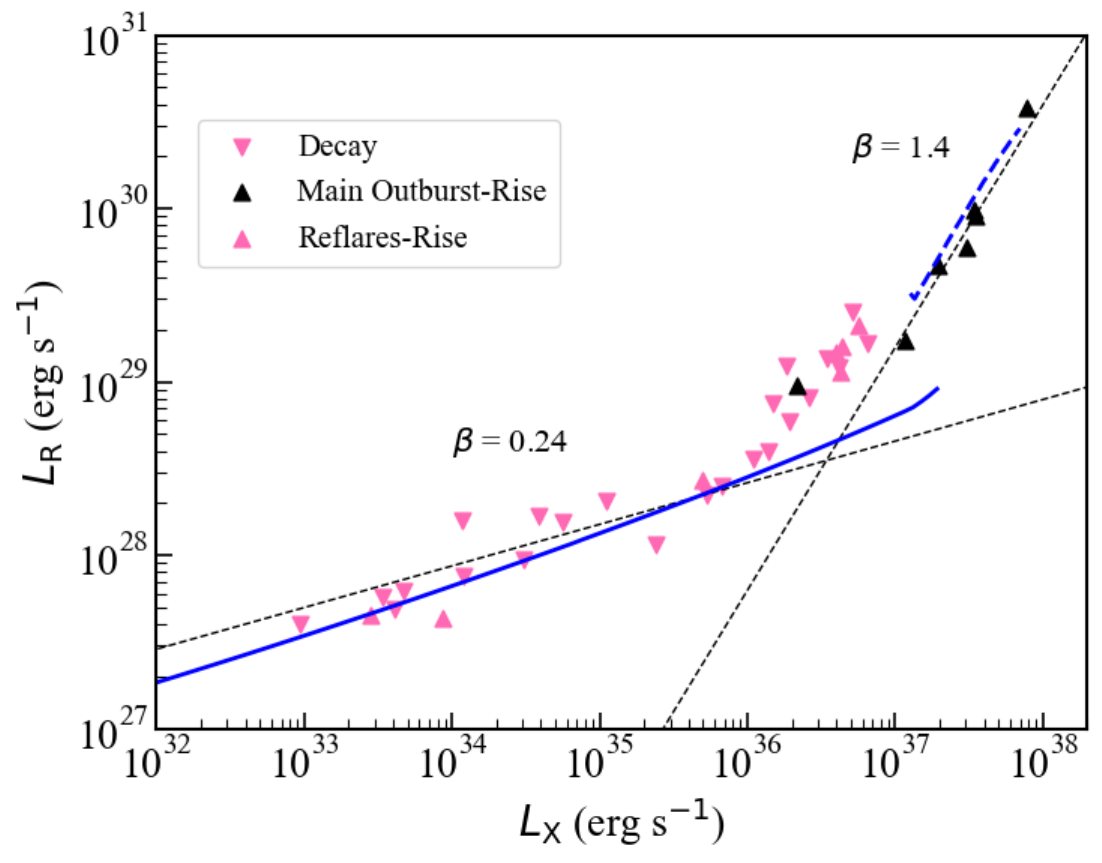}
    \caption{Similar to Figure \ref{fig:5}, but the data points of MAXI J1348-630 are selected. The upward-triangles are plotted with different colors corresponding to different outburst phases, i.e., main outburst (black), and reflares (pink).  The solid blue line represents a soft-to-hard state transition, while the dashed blue line represents a hard-to-soft state transition. The values of parameters: $P_{\rm m}=0.5$, $\alpha_{\rm e} = 0.42$, $\alpha=0.14$, $s=1$, $\Dot{m}_{\rm d} = 0.03$, and $a = 0.6$ are adopted.}
    \label{fig:6}
\end{figure}

\section{Discussion}\label{sec:discussion}

The tight radio/X-ray correlation of BHXRBs in the low/hard state provides important clues on the disc-jet connection. In the low/hard state, the X-ray emission is predominantly from the inner ADAF, of which the radiation efficiency decreases with decreasing dimensionless mass accretion rate $\dot{m}$, because of the radial energy advection \citep[][]{1998tx19.confE.261N,2014ARA&A..52..529Y}. Thus, the hard X-ray luminosity of the BHXRB is governed by the mass accretion rate, which can be calculated with the ADAF spectral calculations \citep[][]{2010ApJ...716.1423X}.

  {We assume that the large-scale magnetic field generated in the outer thin disc is dragged inwards by the inner ADAF as an external field, and it is then substantially enhanced in the region very close to the BH \citep[][]{2021A&A...654A..81C,doi:10.1126/science.abo4504}.} The radio emission is predominantly from the jets, which are supposed to be accelerated by the field co-rotating with the horizon of a spin BH via the Blandford-Znajek mechanism.

The hysteretic state transition indicates that the hard-to-soft state transition is quite different from the soft-to-hard transition, which can be explained in the frame of the accretion disc with magnetically driven outflows \citep[][]{2016ApJ...817...71C,2021A&A...654A..81C}. Based on this scenario,   {we found that the physical processes of the hard-to-soft state and soft-to-hard transitions are quite different} (see the detailed description of the model calculations in Section \ref{subsec:state-transi}). The detailed physics of the accretion mode transition between a thin disc and an ADAF is still unclear, though there are different scenarios suggested in the literature. In this work, we derive the relation of the truncated radius with the accretion rate assuming that an ADAF is always present whenever $\dot{m}(r)<\dot{m}_{\rm cr}(r)$ \citep[][]{1998tx19.confE.261N}.  

Together with the derived relation between the X-ray luminosity and the accretion rate, the correlations of radio/X-ray are available for the soft-to-hard and hard-to-soft state transitions respectively, which are compared with the observations. There are several parameters, $\alpha_{\rm e}$, $s$, $\alpha$, and the BH spin parameter $a$, in our model. We find that the observed normal radio/X-ray correlation (in this work, only the data points observed during soft-to-hard state transitions are used) with a slope of $\beta\simeq 0.6$ can be reproduced quite well by our model calculations for soft-to-hard state transition, when $\alpha_{\rm e}=0.83$ and $s=0.3$ (or $\alpha_{\rm e}=1$ and $s=0.6$), are adopted (see Figures \ref{fig:2} and \ref{fig:3}). The values of $\alpha_{\rm e}$ adopted in the model calculations are roughly consistent with the electron temperature distribution in the outer region of the ADAF \citep[e.g.,][]{2000ApJ...534..734M,2010ApJ...716.1423X}. The radiation efficiency of an ADAF is somewhat quite uncertain, which is affected mainly by the values of the disc parameter $\delta$ and the wind parameter (see the discussion in Section \ref{subsect:l_x}). This issue has been extensively explored in previous works based on hydrodynamic models with a parameterized wind model \citep[e.g.,][]{2012MNRAS.427.1580X}. However, their calculations are different from the ADAF with magnetically driven outflows considered in this work. For simplicity, we assume that the radiation efficiency of the ADAF with magnetic outflows can be described by the parameter $s$ (see the discussion in Section \ref{subsect:l_x}). This issue may be well resolved by a self-consistent calculation of an ADAF with magnetic outflows, which is beyond the scope of this work.

Besides a universal radio/X-ray correlation with $\beta\sim 0.6$, an additional correlation with $\beta\simeq 1.4$ has been observed in some sources with high luminosity \citep[e.g.,][]{2011MNRAS.414..677C,2018MNRAS.481.4513I}, which was interpreted as a sudden increase in radiation efficiency of accretion disc due to the accretion rate exceeding the critical rate on H1743-332  \citep[][]{2011MNRAS.414..677C} and GRS 1915+105 \citep[][]{2010A&A...524A..29R}.This is a long puzzling issue, which invoked a variety of other complicated models to account for \citep[e.g.,][]{2004MNRAS.354..953Y,2009ApJ...699.1919P,2011MNRAS.414..677C,2012MNRAS.427.1580X,2014MNRAS.440..965H,2014A&A...562A.142M,2016MNRAS.456.4377X,2018MNRAS.478.5159M,2019ApJ...871...26K,2019Natur.569..374M,2021MNRAS.504..444C}. 

In this work, we also carry out the calculations of radio/X-ray correlation in the hard-to-soft state transition, and find its slope is much steeper than that observed in the soft-to-hard state transition. The transition luminosity in the hard-to-soft transition is much higher than that in the soft-to-hard state transition (hysteretic state transition), which means the transition accretion rate is high in the hard-to-soft transition. Thus, the outer thin disc is radiation pressure dominated when $\dot{m}_{\rm d}>\dot{m}_{\rm d,rg}$ (see Equation \ref{mdot_pr}). The field generated in the thin disc has a different dependence on the disc properties in the soft-to-hard transition, as the outer thin disc is always gas pressure dominated in the soft-to-hard transition due to lower transition luminosity/accretion rate  (see Equations \ref{b_pd_2} and \ref{b_pd_gas}).  It makes the radio luminosity (jet power) vary more sensitively with the accretion rate, and finally leads to a steeper radio/X-ray correlation in hard-to-soft state transition.

We tentatively compare our model calculations with the observations of an individual source, H1743-332, which has been well observed both in the soft-to-hard and hard-to-soft state transitions \citep[][]{2011MNRAS.414..677C,2018MNRAS.481.4513I}. We first tune the values of the model parameters to fit the radio X-ray correlation of soft-to-hard state transition, and then we find the correlation of the hard-to-soft state transition can be automatically reproduced by the model calculations (see Figure \ref{fig:5} ). It indeed shows a steeper correlation in the hard-to-soft transition with higher luminosity. Our model calculations can naturally reproduce these two correlations between X-ray and radio luminosity, though the values of the parameters adopted are slightly different from those adopted in Figure \ref{fig:2} or \ref{fig:3}.

MAXI J1348-630 is another well-observed source with the hybrid radio/X-ray correlations. During soft-to-hard state transition, it shows a similar track as H1743-322 with a slope $\beta \sim 0.24$ while the radio luminosity is much lower \citep[][]{2021MNRAS.505L..58C}. During the hard-to-soft state transition in the main outburst, the slope of radio/X-ray correlation is close to $1.4$, similar to that observed in H1743-322 (see Figure \ref{fig:6}). It also has multiple re-flares, and the rising phases (hard-to-soft state transitions) of these re-flares are observed, which follow similar radio X-ray correlations as observed in main flares. For H1743-332, there seems lack of sufficient data for radio X-ray correlation analysis during the repeated re-flares.

In both of these two sources, the slope of the radio/X-ray correlation is shallower than $\sim 0.6$ of the correlation observed in soft-to-hard state transition for a sample of BHXRBs. The model parameter, $\alpha_{\rm e}\sim 0.4$, is required in the model fitting on the data of these two sources, which is lower than $\alpha_{\rm e}\sim 0.8-1$ for the normal correlation of BHXRBs. It indicates the physical properties of the ADAF in these two sources may deviate to some extent from most other BHXRBs in the sample. A lower $\alpha_{\rm e}$ means that the electron temperature of the ADAF decreases more slowly radially in these two sources. 
We speculate that the magnetic outflows may significantly alter the ADAF structure, and then the electron temperature distribution, though the detailed physics is still unclear. Some sources even evolve in the radio X-ray plane, from a slope of $\sim 1.4$ to $\sim 0.23$, and then back to the $\sim 0.6$ path as the source decays. It may imply that the properties of the magnetic outflows may evolve drastically with decreasing accretion rates. We know that the disc and outflows are strongly coupled, and a physical disc-outflow model may help us understand this feature.

Our present work focuses on the radio X-ray correlation observed in BHXRBs. One may wonder whether the model can be applied to the neutron star XRBs (NSXRBs). Unfortunately, only the power of the jets magnetically driven by a rotating BH is considered in our model, which is invalid for NSXRBs. Our model can be extended for NSXRBs, if the physics of the jets accelerated by a neutron star is included, which is beyond the scope of this paper.

\section*{Acknowledgements}
We are grateful to the referee for his/her insightful comments and suggestions. This work is supported by the NSFC (12073023, 12233007, 12361131579, 12347103), the science research grants from the China Manned Space Project with No. CMS-CSST- 2021-A06, the fundamental research fund for Chinese central universities (Zhejiang University), and the student research training program (Zhejiang University X2022440). 
B.Y. is supported by NSFC grants 12322307, 12273026; by the National Program on Key Research and Development Project 2021YFA0718500; by the Natural Science Foundation of Hubei Province of China 2022CFB167; by the Fundamental Research Funds for the Central Universities 2042024kf1033; Xiaomi Foundation / Xiaomi Young Talents Program. 
AAZ acknowledges support from the Polish National Science Center under the grants 2019/35/B/ST9/03944, 2023/48/Q/ST9/00138, and from the Copernicus Academy under the grant CBMK/01/24. Thanks to Stephane Corbel (University Paris Diderot) for providing the data of GRS 1915+105.

\section*{Data Availability}

The data underlying this article will be shared on reasonable request to the corresponding author.



\bibliographystyle{mnras}
\bibliography{example} 








\bsp	
\label{lastpage}
\end{document}